\documentclass[aps,pra,reprint,amsmath,amssymb,superscriptaddress,longbibliography]{revtex4-2}

\usepackage{graphicx}
\usepackage{booktabs}
\usepackage{adjustbox}
\usepackage{xcolor}
\usepackage{tikz}
\usetikzlibrary{arrows.meta,positioning}
\usepackage[colorlinks,allcolors=blue!50!black]{hyperref}

\newcommand{\genie}{\textsc{GENIE}}
\newcommand{\kcal}{kcal\,mol$^{-1}$}
\newcommand{\Eh}{E_\mathrm{h}}

\begin{document}

\title{\genie: learned quantum circuits as embedding impurity solvers, and why
compactness, not correlation strength, decides when they work}

\author{Awadhoot Vijay Loharkar}
\email{awadhoot.vijay.loharkar@itwm.fraunhofer.de}
\author{Arcesio Castañeda Medina}
\email{arcesio.castaneda.medina@itwm.fraunhofer.de}
\affiliation{Fraunhofer Institute for Industrial Mathematics ITWM, 67663 Kaiserslautern, Germany}
\date{\today}

\begin{abstract}
Sample-based quantum diagonalization has been combined with density matrix embedding
theory (DMET) to treat molecules larger than a quantum device, but always with
fixed-form circuits, and benchmarked against the embedded exact energy rather than
against classical selection at the same subspace size. We introduce \genie, a DMET
impurity solver in which a generative transformer learns the sampling circuit and the
quantum-selected configuration interaction (QSCI) energy is the training signal. On
N$_2$ and hydrogen chains we compare \genie\ at equal determinant budget with an oracle,
CIPSI, ASCI and heat-bath CI, and with DMET+SQD using an optimized local unitary cluster
Jastrow circuit. \genie\ selects: at 200 determinants (3\,\% of the space) it is within a
factor of 1.7 of CIPSI at every N$_2$ geometry and 25--1000 times more accurate than the
determinants a fixed circuit actually samples, while standard SQD reaches good energies by
filling the space, up to 73\,\% of it for H$_{14}$, still 80--110\,kcal\,mol$^{-1}$ off, and
under realistic noise its subspace becomes the full space. \genie\ does not beat classical
selection: ASCI is 1--5 times better, and on N$_2$/cc-pVDZ CIPSI is 12--23 times better
at \genie's uncapped size. \genie\ fails where the number of determinants carrying 99\,\% of
the state approaches the budget, irrespective of the leading-determinant weight, and the
failures are failures of reach, visible in the selected determinants; a deeper circuit
recovers CIPSI quality for H$_{14}$ at 2.0\,\AA. A CCSD/MP2 amplitude count predicts this
number within a factor of three to four over three orders of magnitude, and a router built
on it never recommended \genie\ where it then failed, apart from seed-to-seed failures; in a
blind test on water and a twelve-atom hydrogen ring it transferred unchanged. We
also identify two benchmarking pitfalls, the orbital basis of the embedding Hamiltonian
and of the classical baseline, that can manufacture apparent advantages of one to two
orders of magnitude.
\end{abstract}

\maketitle

\section{Introduction}\label{sec:intro}

Sample-based quantum diagonalization uses a quantum computer only to propose
determinants: a circuit is prepared and measured, the bitstrings span a subspace, and the
Hamiltonian is diagonalized classically in that subspace. In its original form,
quantum-selected configuration interaction (QSCI)~\cite{kanno2023qsci}, and in its
large-scale form with configuration recovery, sample-based quantum diagonalization
(SQD)~\cite{robledo2025sqd}, the energy is variational and no expectation value has to
be estimated on the device. SQD experiments on up to 77 qubits~\cite{robledo2025sqd} and
its Krylov variants~\cite{skqd2025,sqdrift2025} have made the approach the main route to
chemistry on present hardware.

To treat molecules larger than a device, SQD has been used as the impurity or fragment
solver of quantum embedding: density matrix embedding theory
(DMET)~\cite{knizia2012dmet,knizia2013dmet,wouters2016dmet} for hydrogen rings and
cyclohexane conformers~\cite{shajan2025dmetsqd} and for ligand-like
molecules~\cite{patra2025dmetsqd}, the localized active space (LAS)
method~\cite{hermes2019lasscf,hermes2020fe2bpym} for multi-metal
complexes~\cite{wang2025lassqd}, and DFT embedding~\cite{shiota2026dftqsci}. All of these
sample fixed-form circuits, in practice the local unitary cluster Jastrow (LUCJ)
ansatz~\cite{motta2023lucj}, and report the error against the exact (FCI) energy of the
embedded problem. The only learned component so far is a classical restricted Boltzmann
machine that post-processes the samples~\cite{patra2026rbm}.

An energy close to the embedded FCI does not reveal \emph{why} it is good. It can arise
from \emph{selection}, the circuit sampling the important determinants, or from
\emph{filling}, the subspace becoming a large fraction of the space. Classical selected
configuration interaction (SCI), such as CIPSI~\cite{huron1973cipsi}, adaptive sampling CI
(ASCI)~\cite{tubman2016asci} and heat-bath CI (HCI)~\cite{holmes2016hci}, selects
deterministically and is the natural yardstick. Critical assessments found QSCI expansions
less compact than classical SCI and sampling inefficient for
peaked states~\cite{reinholdt2025critical}, and a recent review places most claimed
QSCI advantages within the classically simulable regime~\cite{mlsqdreview2026}; strong
classical SCI~\cite{zhang2025trimci} raises the bar further, while advantages for sparse
ground states have been shown for constructed Hamiltonians~\cite{kirby2026sparse}.

A different line of work makes the circuit itself learnable. The generative quantum
eigensolver (GQE)~\cite{nakaji2024gqe} trains a transformer to emit sequences of
operators from a pool. Trained on the QSCI energy, it produces subspaces up to 50\,\%
smaller than HCI for N$_2$ in active spaces of up to 32 qubits with 83--98\,\% fewer
two-qubit gates than LUCJ~\cite{kemmoku2026gqeqsci}, and has been scaled to 40
qubits~\cite{elgammal2026gqe40}. These studies work in a complete active space, without
embedding, and compare against HCI only.

Here we combine the two lines and ask a sharper question than ``is the energy good'':
\emph{does the circuit select, how does it compare with the best classical heuristics at
the same number of determinants, and can one tell beforehand whether it will work?} Our
contributions are:
(i)~\genie, a DMET impurity solver whose sampling circuit is learned by GQE, with the
embedding Hamiltonian expressed in the embedded mean-field basis so that operator pool,
reference determinant and Hamiltonian are consistent (Sec.~\ref{sec:methods});
(ii)~an equal-budget protocol that compares every \genie\ run at its final subspace size
$N$ with an oracle, CIPSI, ASCI and HCI in two orbital bases, and with DMET+SQD restricted
to the determinants actually sampled (Secs.~\ref{sec:res-budget}--\ref{sec:res-sqd});
(iii)~the finding that failures are governed by the compactness of the impurity
wavefunction relative to the budget rather than by correlation strength
(Sec.~\ref{sec:res-compact}); and (iv)~an amplitude-based estimate and router that
predict these failures in advance, validated in a blind test on new systems
(Secs.~\ref{sec:res-router} and \ref{sec:res-blind}). Section~\ref{sec:res-routes}
discusses the scope of an impurity-level study and the LAS alternative, and Sec.~\ref{sec:res-cost}
resources and hardware feasibility; the appendices contain
every data table behind the figures.

\section{Methods}\label{sec:methods}

\subsection{Density matrix embedding}\label{sec:m-dmet}
The whole molecule is treated at the unrestricted Hartree--Fock (UHF) level. The impurity
orbitals are the active orbitals chosen with atomic valence active space (AVAS) projection~\cite{sayfutyarova2017avas}
onto atom-resolved labels (e.g.\ the 1s functions of the central atoms of a hydrogen
chain). A Schmidt decomposition of the reference one-particle density matrix gives a bath
of at most the same size; impurity plus bath span the embedding space of $n$ orbitals,
in which the interacting Hamiltonian is
\begin{align}
H_\mathrm{emb}={}&E_\mathrm{core}+\sum_{pq\sigma}h_{pq}\,a^\dagger_{p\sigma}a_{q\sigma}\nonumber\\
&+\tfrac12\sum_{pqrs\sigma\tau}(pq|rs)\,a^\dagger_{p\sigma}a^\dagger_{r\tau}a_{s\tau}a_{q\sigma},
\label{eq:hemb}
\end{align}
with the frozen environment folded into $E_\mathrm{core}$ and
$h_{pq}$~\cite{knizia2012dmet,knizia2013dmet,wouters2016dmet}. Unlike the textbook
mean-field bath, the Schmidt decomposition acts on a correlated reference density, a CASCI
in the AVAS space, and the bath size is chosen adaptively from the singular-value spectrum;
the scheme is one-shot, without a correlation potential (App.~\ref{app:dmet}). An embedded
RHF is solved in the embedding space with $E_\mathrm{core}$ as constant.

\paragraph*{Orbital basis.} Equation~\eqref{eq:hemb} is naturally expressed in the raw
(Schmidt) basis. We rotate it into the embedded RHF molecular-orbital (MO) basis,
$h\to C^\mathsf{T}hC$ and likewise for $(pq|rs)$. Energies are invariant, but this step is
essential for a learned circuit: the circuit starts from ``the lowest orbitals occupied''
and its excitation operators are indexed by occupied and virtual MOs. In the raw basis the
starting determinant lies 20\,000--40\,000\,\kcal\ above RHF and \genie\ errors were
40--95\,\kcal; after rotation they fall by up to five orders of magnitude
(App.~\ref{app:basis}, Table~\ref{tab:basis}). A complete-active-space GQE pipeline is
unaffected, because its basis already is the MO basis, which is why the problem is easy to
miss when such a pipeline is moved into an embedding.

\subsection{Quantum-selected configuration interaction}\label{sec:m-qsci}
A state $|\psi\rangle$ prepared on $2n$ qubits (Jordan--Wigner) is measured $N_s$ times.
After discarding bitstrings with the wrong electron numbers and completing the set under
$\alpha\leftrightarrow\beta$ exchange (spin completion), the distinct determinants form
$\mathcal{S}$, $|\mathcal{S}|\le N$, where $N$ is the budget (cap). The QSCI energy is the
lowest eigenvalue of the projected Hamiltonian,
\begin{equation}
E_\mathcal{S}=\min\operatorname{eig}\,\bigl[\langle D_I|H_\mathrm{emb}|D_J\rangle\bigr]_{I,J\in\mathcal{S}},
\label{eq:qsci}
\end{equation}
which is variational, $E_\mathcal{S}\ge E_\mathrm{FCI}^\mathrm{emb}$. SQD~\cite{robledo2025sqd}
instead diagonalizes in the Cartesian product of all sampled $\alpha$ strings with all
sampled $\beta$ strings (``product'' subspace) and recovers configurations broken by
noise; we also evaluate the ``list'' subspace of the sampled $(\alpha,\beta)$ pairs only,
which isolates what the circuit selects.

\subsection{\genie: generative circuits for the impurity}\label{sec:m-genie}
\begin{figure*}[t]
\centering
\includegraphics[width=\textwidth]{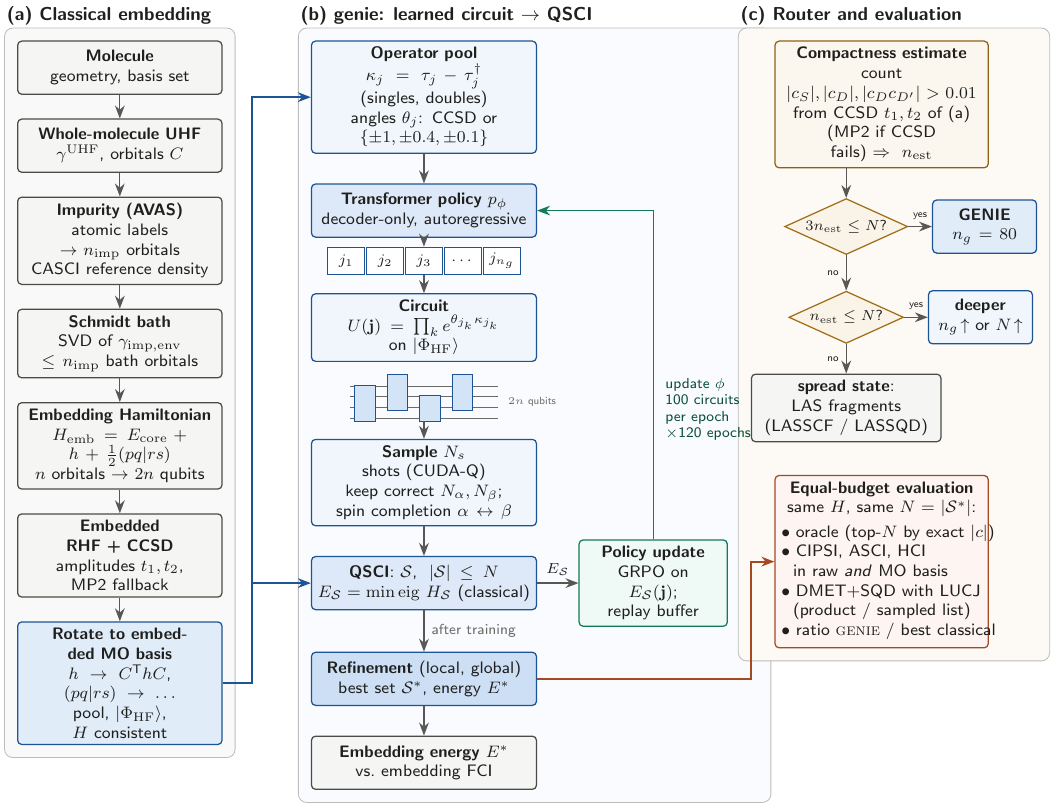}
\caption{The \genie\ pipeline. (a)~Classical embedding: an AVAS impurity and its Schmidt
bath define the embedding Hamiltonian $H_\mathrm{emb}$ of $n$ orbitals ($2n$ qubits); the
embedded RHF provides CCSD amplitudes (operator angles and compactness estimate) and the
orbitals into which $H_\mathrm{emb}$ is rotated, so that operator pool, reference
determinant and Hamiltonian share one basis. (b)~Learned circuit: a transformer emits a
sequence of $n_g$ excitation operators; the circuit is sampled, the determinants are
filtered and spin-completed, and the QSCI energy in at most $N$ determinants is the
signal that updates the transformer. After training, the best determinant set is refined,
and its energy is compared with the exact solution of the embedding. (c)~Before training, the router compares the
estimated compactness $n_\mathrm{est}$ with the budget $N$; after training, every result is
compared with classical selection and with DMET+SQD at the same $N$ on the same
Hamiltonian.}
\label{fig:workflow}
\end{figure*}

\paragraph*{Circuit generation.} The operator pool consists of fermionic single and double
excitation generators $\kappa_j=\tau_j-\tau_j^\dagger$ of the embedded RHF reference,
each with a fixed angle $\theta_j$, taken either from the CCSD amplitudes of the embedded
RHF (``CCSD angles'') or from the set $\{\pm1,\pm0.4,\pm0.1\}$ (``free angles''). A
decoder-only transformer generates a sequence $\mathbf{j}=(j_1,\dots,j_{n_g})$
autoregressively, and the circuit is
\begin{equation}
U(\mathbf{j})=\prod_{k=n_g}^{1}\exp\bigl(\theta_{j_k}\kappa_{j_k}\bigr),\qquad
|\psi(\mathbf{j})\rangle=U(\mathbf{j})|\Phi_\mathrm{HF}\rangle .
\label{eq:circuit}
\end{equation}

\paragraph*{Reward and policy update.} Each generated circuit is sampled, the QSCI
energy $E_\mathcal{S}(\mathbf{j})$ of Eq.~\eqref{eq:qsci} is computed classically, and the
transformer is updated with a group-relative policy-gradient (GRPO) objective, as in the
GQE-for-QSCI trainer~\cite{kemmoku2026gqeqsci}, rather than the logit-matching loss of the
original GQE~\cite{nakaji2024gqe}. Token probabilities are $\pi_\theta(j)=\mathrm{softmax}
(-\beta z_j)$ for transformer outputs $z_j$ and an inverse temperature $\beta$. For a batch of
$B$ sequences with energies $E_b$, the advantages are $A_b=(\bar E-E_b)/\sigma_E$ (batch mean
and standard deviation), and with probability ratios $r_{bk}=\pi_\theta(j_{bk})/
\pi_{\theta_\mathrm{old}}(j_{bk})$ the loss is
\begin{multline}
\mathcal L=-\frac{1}{n_g}\sum_{k}\log\pi_\theta(j_{b^\ast k})\\
-\min\Bigl(\bigl\langle r_{bk}A_b\bigr\rangle,\ \bigl\langle
\mathrm{clip}(r_{bk},1-\epsilon_\mathrm{l},1+\epsilon_\mathrm{h})\,A_b\bigr\rangle\Bigr),
\label{eq:loss}
\end{multline}
where $b^\ast$ is the lowest-energy sequence of the batch, $\langle\cdot\rangle$ averages over
sequences and positions, and $\epsilon_\mathrm{l}=0.2$, $\epsilon_\mathrm{h}=0.28$. The first
term reinforces the best circuit found so far in the batch; the second is a clipped
(PPO-style) surrogate. $\beta$ starts at 0.5 and is raised in steps of 0.02 by a scheduler
that targets an energy variance of $10^{-5}$ within a batch, so that sampling sharpens as the
batch converges. Optimizer: AdamW (learning rate $5\times10^{-6}$, weight decay 0.01).
A replay buffer keeps the best sequences.

\paragraph*{Refinement.} As in Ref.~\cite{kemmoku2026gqeqsci}, the QSCI solutions of the
circuits of one epoch are merged into one sparse wavefunction (local refinement), which is
then merged with the result of all previous epochs (global refinement). Each merge solves
the generalized eigenvalue problem in the non-orthogonal basis of the input wavefunctions
and keeps the $N$ determinants with the largest weights. Refinement is purely classical and
generates no new determinants: the final set $\mathcal S^\ast$ consists of determinants
sampled by some circuit during training. It is therefore the product of all circuits
executed in training, not of a single circuit, for H$_{14}$ at 2.0\,\AA\ the best single
circuit samples 1\,400--3\,400 distinct determinants, the refined set holds $10^4$, and the
quantum cost of a \genie\ point is that of the whole training run
(Sec.~\ref{sec:res-cost}).

Unless stated, $n_g=80$, 120
epochs, 100 circuits per epoch (12\,000 circuits per run, $N_\mathrm{s}=10^5$ shots each, i.e.\ $1.2\times10^9$ shots), a QSCI cap of $N=10^4$ (or budgets of 100 and 200), and
three seeds (App.~\ref{app:comp}, Table~\ref{tab:hyper}); runs that stop before 120
epochs are rejected by the pipeline. All simulations are noiseless state-vector
simulations with CUDA-Q~\cite{cudaq}. Because QSCI energies are variational, the best of
$k$ seeds is a valid result at $k$-fold cost.

\subsection{Equal-budget protocol and baselines}\label{sec:m-baselines}
For every finished \genie\ run with final subspace size $N$ we evaluate, on the same
$H_\mathrm{emb}$ and at the same $N$, the variational energies of
(a)~an \emph{oracle}, the $N$ determinants with largest $|c_I|$ in the exact embedded
ground state, re-diagonalized (a reference, not a method);
(b)~\emph{CIPSI}~\cite{huron1973cipsi}, grown by Epstein--Nesbet second-order
contributions until exactly $N$ determinants;
(c)~\emph{ASCI}~\cite{tubman2016asci}, fixed-size and seeded with CIPSI, which can drop
determinants; and (d)~a fixed-size \emph{HCI}~\cite{holmes2016hci} (App.~\ref{app:baselines}).
Selection is basis dependent, so each selector is run in both the raw and the MO basis and
the better value is the reference. Errors are quoted against the embedding FCI.

For DMET+SQD we use the same embedding and an LUCJ circuit~\cite{motta2023lucj} with
all-pairs interactions and two repetitions, CCSD-initialized and optimized with the linear
method of ffsim~\cite{ffsim}, $10^4$ or $10^5$ shots, and six configuration-recovery
iterations~\cite{robledo2025sqd} (Qiskit addon~\cite{qiskit2024}). Noise is modeled on the
FakeSherbrooke backend (App.~\ref{app:noise}).

\subsection{Compactness and router}\label{sec:m-router}
We characterize the exact impurity state $\sum_I c_I|D_I\rangle$ by its leading weight
$w_1=\max_I|c_I|^2$ and by its compactness $n_{99}$, the smallest number of determinants,
ordered by $|c_I|$, whose weights sum to $0.99$. Since $n_{99}$ requires the exact state,
we estimate it from CCSD amplitudes (MP2 amplitudes if CCSD does not converge). With
determinant-level single and double coefficients $c_S$ and $c_D$ (from $t_1$ and
$\tau_2=t_2+t_1t_1$; App.~\ref{app:router}) and disconnected quadruples approximated by
products,
\begin{equation}
n_\mathrm{est}=1+\#\{|c_S|>\tau\}+\#\{|c_D|>\tau\}+\#\{|c_Dc_{D'}|>\tau\},
\label{eq:nest}
\end{equation}
with $\tau=0.01$. The router recommends \genie\ if $3n_\mathrm{est}\le N$; \genie\ with a
deeper circuit or a larger budget if $n_\mathrm{est}\le N<3n_\mathrm{est}$; and a fragment
method (LAS) otherwise. The factor three is the observed accuracy of $n_\mathrm{est}$
(Sec.~\ref{sec:res-router}).

\section{Numerical results}\label{sec:results}
We study N$_2$/STO-3G at $R=1.1$--$2.5$\,\AA, where the AVAS N\,2s2p impurity plus bath
spans the full valence space (9--10 orbitals, 18--20 qubits), so that the embedding error
vanishes and the solver is isolated; N$_2$/cc-pVDZ (13 of 28 orbitals, 26 qubits,
$2.9\times10^6$ determinants); linear H$_{10}$ and H$_{14}$/STO-3G with central-atom
impurities at $R=1.0$, 1.5 and 2.0\,\AA\ (up to 20 qubits, 63\,504 determinants); and
[Fe(H$_2$O)$_4$]$_2$bpym$^{4+}$ for the LAS route~\cite{wang2025lassqd,hermes2020fe2bpym}.
All systems and embeddings are listed in Table~\ref{tab:systems}. Software: PySCF~\cite{sun2020pyscf},
CUDA-Q~\cite{cudaq}, Qiskit~\cite{qiskit2024}, ffsim~\cite{ffsim} and our package
QuEnAIS~\cite{quenais}.

\subsection{Selection at equal budget}\label{sec:res-budget}
\begin{figure}[t]
\centering
\includegraphics[width=\columnwidth]{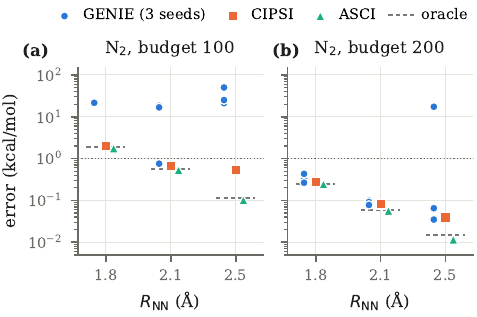}
\caption{N$_2$/STO-3G, error against the embedding FCI at fixed budgets of (a) 100 and
(b) 200 determinants: \genie\ (CCSD angles, 80 operators, three seeds), CIPSI and ASCI (MO
basis), and the oracle (dashed). Dotted line: 1\,\kcal. Full data in
Table~\ref{tab:n2full}.}
\label{fig:budget}
\end{figure}
Figure~\ref{fig:budget} shows the central comparison. At 200 determinants (about 3\,\% of
the space) \genie\ is within a factor of 1.7 of CIPSI at every geometry
[Fig.~\ref{fig:budget}(b)], including 2.5\,\AA\ where the leading determinant carries only
10\,\% of the weight. ASCI is 1--5 times better than both and coincides with the oracle:
\genie\ reaches CIPSI but not ASCI quality. One 200-determinant seed in three fails
(17\,\kcal\ at 2.5\,\AA). At 100 determinants \genie\ fails at 1.8--2.5\,\AA\
[Fig.~\ref{fig:budget}(a)], with errors of 17--50\,\kcal\ against 0.5--2\,\kcal\ for
CIPSI; only one seed at 2.1\,\AA\ succeeds. Without a cap, \genie\ keeps 600--1\,100
determinants and reaches $10^{-2}$--$10^{-4}$\,\kcal; with free angles it reaches the
embedding FCI using 25--51\,\% of the space.

\subsection{Two benchmarking pitfalls, and seed variance}\label{sec:res-pitfalls}
\begin{figure*}[t]
\centering
\includegraphics[width=\textwidth]{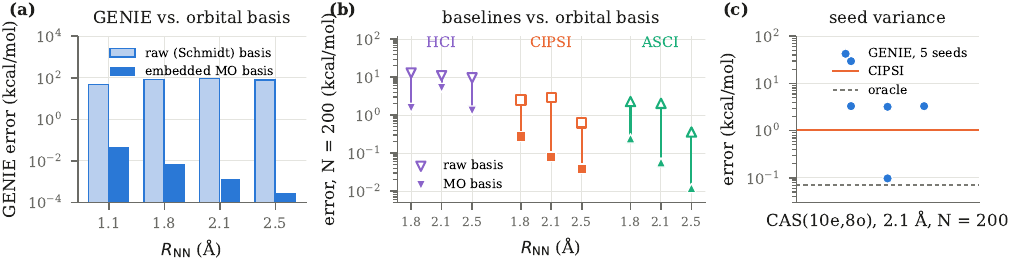}
\caption{(a) \genie\ error for N$_2$/STO-3G with the embedding Hamiltonian in the raw
Schmidt basis and in the embedded MO basis (no cap, CCSD angles, 80 operators;
Table~\ref{tab:basis}). (b) Classical selectors at $N=200$, selected in the raw basis (open)
and in the MO basis (filled); same Hamiltonian, energies are basis invariant, only the
selection changes. (c) Five \genie\ seeds in a plain CAS(10e,8o) N$_2$ problem at 2.1\,\AA,
$N=200$, against CIPSI and the oracle.}
\label{fig:pitfalls}
\end{figure*}
Three effects can make a quantum selection scheme look much better or worse than it is
(Fig.~\ref{fig:pitfalls}). First, the basis of the Hamiltonian: in the raw Schmidt basis
the learned circuit starts from a determinant tens of thousands of \kcal\ above RHF and
fails by 46--95\,\kcal; in the MO basis the same pipeline reaches $10^{-2}$--$10^{-4}$\,\kcal\
[Fig.~\ref{fig:pitfalls}(a)]. Second, the basis of the \emph{baseline}: CIPSI selected in the
raw basis is 2--36 times worse than in the MO basis, HCI and ASCI behave alike
[Fig.~\ref{fig:pitfalls}(b)]. Against raw-basis CIPSI, \genie\ appeared 6--38 times better,
an advantage that disappears under a fair baseline. Our fixed-size HCI is moreover 3--65
times worse than CIPSI on the same problems (Table~\ref{tab:n2full}). This fixed-size
variant is weaker than production semistochastic HCI, which we did not run; the point is
that HCI's quality depends on its implementation and size control, so advantages measured against HCI
alone~\cite{kemmoku2026gqeqsci,weaving2025compact} should be read with care.

Third, seed variance is large [Fig.~\ref{fig:pitfalls}(c)]: five seeds span a factor of
300; the best run beats CIPSI by $10\times$, the median is $3\times$ worse and one in five
fails (App.~\ref{app:baselines}). A single favorable run is therefore not evidence of an
advantage; we report all seeds, and since QSCI energies are variational, the best of $k$
seeds is a legitimate result at $k$-fold cost.

\subsection{Selection versus filling}\label{sec:res-sqd}
\begin{figure}[t]
\centering
\includegraphics[width=\columnwidth]{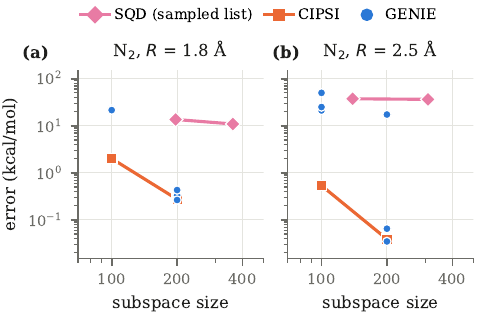}
\caption{N$_2$/STO-3G at (a) 1.8 and (b) 2.5\,\AA: error versus subspace size for the
determinants actually sampled by DMET+SQD (``list'' subspace, optimized LUCJ, $10^4$ and
$10^5$ shots), \genie\ (budgets 100 and 200) and CIPSI. Full data in
Table~\ref{tab:sqdlist}.}
\label{fig:sqd}
\end{figure}
Restricted to the determinants it samples, DMET+SQD is 1.4--100 times less accurate than
CIPSI at equal size (Table~\ref{tab:sqdlist}), and more shots do not help at 2.5\,\AA:
37.4 and 36.4\,\kcal\ for 139 and 309 determinants. The optimized fixed-form circuit does
not reach the important determinants of a stretched bond. With fewer determinants
\genie\ is 25--50 times (1.8\,\AA) and 500--1000 times (2.5\,\AA) more accurate
(Fig.~\ref{fig:sqd}), i.e.\ it behaves like CIPSI rather than like the sampler.

\begin{figure*}[t]
\centering
\includegraphics[width=\textwidth]{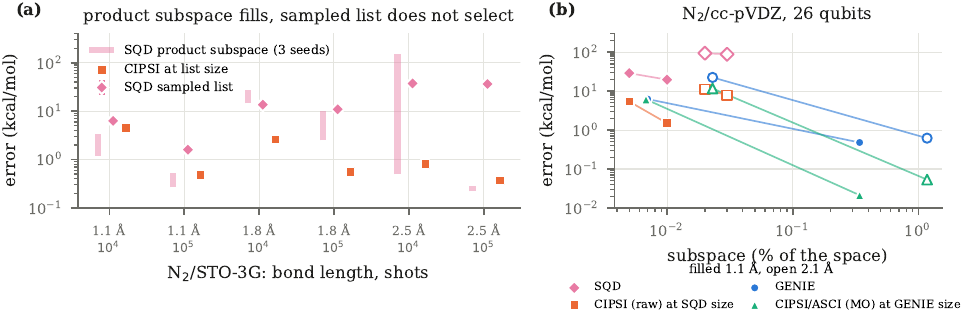}
\caption{(a) N$_2$/STO-3G, DMET+SQD with optimized LUCJ: range over three seeds of the
product-subspace energy (bars), the sampled-list energy (diamonds, mean and range) and
CIPSI at the size of the list (squares). (b) N$_2$/cc-pVDZ (26 qubits): error versus subspace
fraction for DMET+SQD (unoptimized LUCJ, $10^4$/$10^5$ shots), CIPSI (raw basis) at the
same sizes, \genie\ (budget 200 and uncapped), and the best of CIPSI and ASCI in the MO basis at
\genie's sizes; filled 1.1\,\AA, open 2.1\,\AA.}
\label{fig:filling}
\end{figure*}
The good energies usually reported for SQD come from the product subspace. In small
embeddings the $\alpha\times\beta$ product of the sampled strings covers most of the
space; adding random exploration shots made every tested system exact, including
N$_2$/cc-pVDZ, by covering 99.9--100\,\% of spaces of up to $2.9\times10^6$ determinants
(App.~\ref{app:sqd}). This is exact diagonalization, not compression. Per seed, the product
energy scatters over up to two orders of magnitude while the list energy is stable and far
from CIPSI [Fig.~\ref{fig:filling}(a)]. Noise makes this explicit
(App.~\ref{app:noise}, Table~\ref{tab:noise}). Under a FakeSherbrooke noise model (N$_2$,
1.1\,\AA; routed LUCJ with 3\,499 CX gates, estimated circuit fidelity 0.05 at one tenth of
present-day error rates and $10^{-13}$ at present-day rates), the product subspace becomes
the full space and its energy exact. The sampled list \emph{improves} with noise: from
5.5--6.7\,\kcal\ with 87--95 determinants without noise to 0.05--0.20\,\kcal\ with
5\,600 determinants (39\,\% of the space) at one tenth of present-day rates, and
0.2--2.0\,\kcal\ with 7\,600 (53\,\%) at present-day rates. Better energies from a
noisier device are the signature of filling: noise spreads the samples over the space,
and the classical diagonalization does the rest.

On N$_2$/cc-pVDZ (26 qubits, $2.9\times10^6$ and $8.5\times10^5$ determinants at 1.1 and
2.1\,\AA), \genie\ is again far ahead of DMET+SQD at similar size, 6.2 against 29\,\kcal\
and 22.5 against 94\,\kcal\ at about 200 determinants, and without cap reaches 0.48 and
0.62\,\kcal\ with 0.3 and 1.2\,\% of the space [Fig.~\ref{fig:filling}(b),
Table~\ref{tab:dz}]. Against classical selection in the MO basis at the same sizes it is
behind, however: tied with CIPSI and ASCI at 200 determinants and 1.1\,\AA\ (6.25 vs.\ 6.14),
1.9 times behind at 2.1\,\AA\ (22.5 vs.\ 11.7), and 23 and 12 times behind uncapped
(0.48 vs.\ 0.021 and 0.62 vs.\ 0.053\,\kcal). In the larger space, the 80-operator circuit
reaches chemical accuracy but not the compact expansion that CIPSI finds.

\subsection{A genuine embedding: hydrogen chains and circuit depth}\label{sec:res-h14}
\begin{table}[t]
\caption{H$_{14}$ (DMET, 18--20 qubits): \genie\ against classical selection at equal size
$N$ (\kcal\ vs.\ embedding FCI, MO basis). $w_1$, $n_{99}$: leading weight and number of
determinants holding 99\,\% of the exact embedded state (MO basis); $n_g$: operators per circuit. Full
data in Table~\ref{tab:h14full}.}
\label{tab:h14}
\begin{adjustbox}{max width=\columnwidth}
\begin{tabular}{lrrrrrrr}
\toprule
$R$ (\AA) & $w_1$ & $n_{99}$ & $N$ & $n_g$ & \genie & CIPSI & ASCI\\
\midrule
1.0 & 0.89 & 112    & 200    & 80  & 4.60 & 4.50 & 4.50\\
1.5 & 0.46 & 4\,769 & $10^4$ & 80  & 3.0--3.2 & 2.55 & 2.50\\
1.5 & 0.46 & 4\,769 & $10^4$ & 160 & 3.03 & 2.55 & 2.50\\
2.0 & 0.14 & 13\,376 & $10^4$ & 80  & 68--70 & 7.00 & 5.95\\
2.0 & 0.14 & 13\,376 & $10^4$ & 160 & 7.92 & 7.00 & 5.95\\
2.0 & 0.14 & 13\,376 & $10^4$ & 320 & 8.09 & 7.00 & 5.95\\
\bottomrule
\end{tabular}
\end{adjustbox}
\end{table}
\begin{figure}[t]
\centering
\includegraphics[width=\columnwidth]{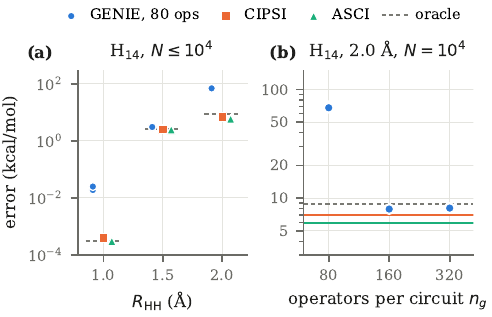}
\caption{H$_{14}$/STO-3G, DMET with a four-atom impurity. (a) Error at $N\le10^4$
versus bond length: \genie\ (80 operators, CCSD and free angles, two seeds each), CIPSI,
ASCI and oracle at the same $N$. (b) 2.0\,\AA: \genie\ error versus circuit length $n_g$;
horizontal lines as in (a). All runs 120 epochs.}
\label{fig:h14}
\end{figure}
\begin{figure*}[t]
\centering
\includegraphics[width=0.8\textwidth]{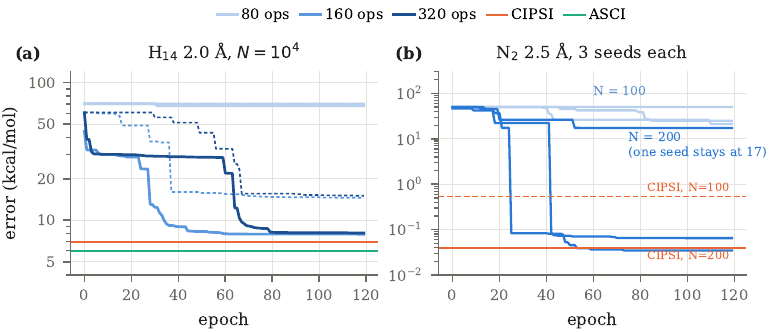}
\caption{Training curves: best error so far against epoch. (a) H$_{14}$ at 2.0\,\AA,
$N=10^4$, with 80 (two seeds), 160 and 320 operators; solid after refinement, dashed for the
raw circuit's subspace; horizontal lines: CIPSI and ASCI at the same $N$. (b) N$_2$ at
2.5\,\AA, three seeds each at $N=100$ (light) and $N=200$ (dark), with CIPSI at both sizes.}
\label{fig:training}
\end{figure*}
For H$_{14}$, DMET truncates the system to 9--10 of 14 orbitals and the embedding is
genuine (Table~\ref{tab:h14}, Fig.~\ref{fig:h14}). Near equilibrium \genie\ ties with CIPSI at 200
determinants, and with $N\approx7\,000$ free-angle circuits reach the embedding FCI. At
1.5\,\AA\ it is 1.2 times behind CIPSI at $10^4$ determinants. At 2.0\,\AA\ and 80
operators it fails: 68--70\,\kcal\ against 6--7\,\kcal\ for CIPSI and ASCI at the same
$10^4$ determinants, identically for both seeds and both angle schemes. At 1.5 and
2.0\,\AA\ every 80-operator run fills the cap of $10^4$ determinants (15.7\,\% of the
space), so the limit is not the number of determinants but which ones the circuit
reaches. Doubling the circuit to 160 operators recovers CIPSI quality [Fig.~\ref{fig:h14}(b)]:
7.92\,\kcal\ against 7.00 for CIPSI; 320 operators add nothing (8.09\,\kcal). At 1.5\,\AA, where
the 80-operator circuit already reaches the important determinants, 160 operators change
nothing (3.03 against 3.04--3.19\,\kcal): depth helps only where reach is the limit.

The training curves (Fig.~\ref{fig:training}) show how. With 80 operators the best energy
stops improving after a few epochs. With 160 and 320 operators it drops in steps, as the
transformer finds sequences that reach new groups of determinants, and settles after
50--70 epochs. Refinement matters: the raw circuit's subspace gives 14.6\,\kcal\ at
160 operators, the refined set 7.9\,\kcal. For N$_2$ at 2.5\,\AA\ the same curves show the seed
failures of Sec.~\ref{sec:res-pitfalls}: two of three seeds at $N=200$ jump to below
0.1\,\kcal\ within 25--45 epochs, the third never does [Fig.~\ref{fig:training}(b)].

H$_{10}$ embeddings are small
($\le12$ qubits) and every method is exact (Table~\ref{tab:diag}).

DMET+SQD on the same H$_{14}$ embeddings ($10^4$ shots, three seeds; Table~\ref{tab:h14sqd})
repeats the N$_2$ picture at larger scale. Its sampled lists of 240--4\,100 determinants
are 5.4--6.1, 78.4--78.6 and 122--128\,\kcal\ off at 1.0, 1.5 and 2.0\,\AA; its product
subspaces are 2.3--2.8, 75--94 and 82--109\,\kcal\ off although they hold 18--25\,\%,
10--12\,\% and 66--73\,\% of the space. \genie\ with $10^4$ determinants (15.7\,\%) reaches
3.0--3.2\,\kcal\ at 1.5\,\AA\ and, with 160 operators, 7.9\,\kcal\ at 2.0\,\AA.

\subsection{Compactness, not correlation strength}\label{sec:res-compact}
\begin{figure*}[t]
\centering
\includegraphics[width=0.8\textwidth]{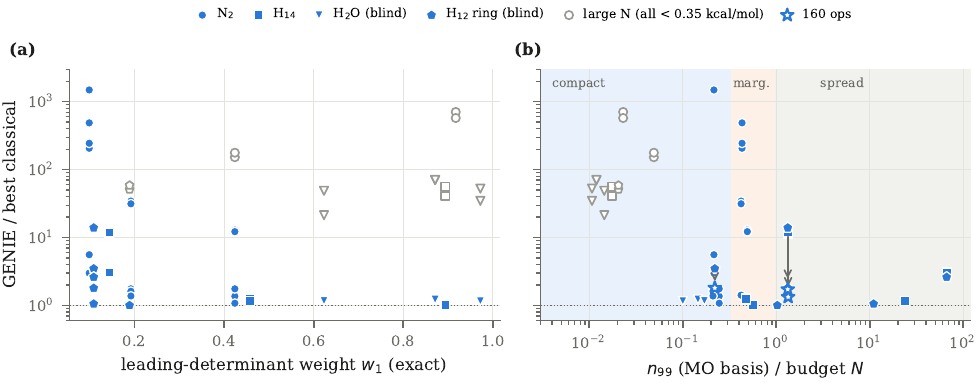}
\caption{\genie\ error relative to the best classical selector at the same size against
(a) the exact leading weight $w_1$ and (b) the compactness $n_{99}$ relative to the budget
$N$ in the MO basis (80 operators; stars and arrows: the same runs with 160 operators, H$_{14}$ at 2.0\,\AA\ and
the H$_{12}$ ring at 1.5 and 2.0\,\AA).
Filled: budgeted runs of N$_2$ and H$_{14}$ and of the blind test (H$_2$O, H$_{12}$ ring,
Sec.~\ref{sec:res-blind}). Open grey: large-$N$ runs, where \genie\ is below
0.35\,\kcal\ and CIPSI nearly exact. Correlation strength does not order the failures;
compactness does.}
\label{fig:compact}
\end{figure*}
\begin{figure*}[t]
\centering
\includegraphics[width=\textwidth]{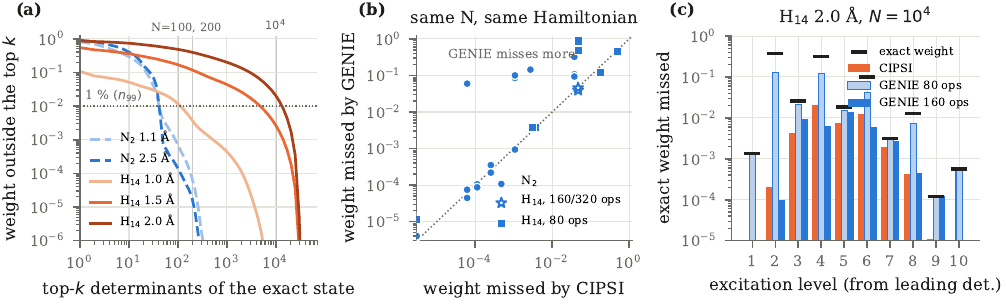}
\caption{What the circuit selects. (a) Weight of the exact embedded state outside its $k$
largest determinants (MO basis); the crossing with 1\,\% defines $n_{99}$. Vertical lines:
budgets used. (b) Exact weight missed by the \genie\ determinant set against that missed by
CIPSI at the same $N$, for every N$_2$ and H$_{14}$ run (H$_{10}$ and runs with both below
$10^{-6}$ omitted). (c) H$_{14}$ at 2.0\,\AA, $N=10^4$: exact weight per excitation level
relative to the leading determinant (bars: weight missed; dashes: total weight at that
level) for CIPSI and \genie\ with 80 and 160 operators.}
\label{fig:anatomy}
\end{figure*}

The leading weight does not order these outcomes [Fig.~\ref{fig:compact}(a)]: N$_2$ at
2.5\,\AA\ ($w_1=0.10$) is handled, H$_{14}$ at 2.0\,\AA\ ($w_1=0.14$) is not. What differs is
how many determinants carry the state [Fig.~\ref{fig:anatomy}(a)]: 42--49 for stretched
N$_2$, where three bonds break between two atoms, but 13\,376 for H$_{14}$ at 2.0\,\AA, where
every embedded bond breaks (Table~\ref{tab:diag}; all counts in the MO basis in which
\genie\ works). Against $n_{99}/N$ the picture is ordered [Fig.~\ref{fig:compact}(b)]. Below
$n_{99}/N\approx1/3$, \genie\ stays within a small factor of the best classical selector,
apart from failed seeds. Between 1/3 and 1 the outcome depends on the circuit: N$_2$ at
$N=100$ ($n_{99}/N=0.42$--0.49) fails, while H$_{14}$ at 1.0\,\AA\ ($N=200$, 0.56) and
1.5\,\AA\ ($N=10^4$, 0.48) succeed. Above 1, every selected-CI method fails in absolute
terms, because the budget is smaller than the state; there, H$_{14}$ at 2.0\,\AA\ needs 160
operators to keep up with CIPSI, and so does the H$_{12}$ ring at 2.0\,\AA\
(Sec.~\ref{sec:res-blind}).

The determinant sets themselves show what goes wrong [Fig.~\ref{fig:anatomy}(b,c)]. Where
\genie\ matches CIPSI in energy, it also captures the same exact weight, typically missing
$10^{-4}$--$4\times10^{-3}$ of it. The failures are failures of reach: at $N=100$ the N$_2$ sets miss
9--32\,\% of the weight against 0.1--4\,\% for CIPSI, and for H$_{14}$ at 2.0\,\AA\ the
80-operator circuits miss 48\,\% (CCSD angles) and 85--88\,\% (free angles) against 4.7\,\%
for CIPSI. Resolved by excitation level, the 80-operator set misses 13\,\% of the weight in
doubles and 12\,\% in quadruples of the leading determinant, which an 80-operator product of
excitations cannot generate in sufficient number. With 160 operators the missed
weight drops to 3.9\,\%, below CIPSI's 4.7\,\%; the remaining loss sits in odd excitation
levels, which CIPSI covers better [Fig.~\ref{fig:anatomy}(c)].

Compactness is basis dependent. In the raw Schmidt basis, which is localized,
$n_{99}$ of H$_{14}$ is 1.3--1.8 times \emph{smaller} (87, 2\,981 and 7\,648 at 1.0, 1.5 and 2.0\,\AA),
whereas for N$_2$ it is 1.3--2.2 times larger (64--91; Table~\ref{tab:diag}). A stretched
hydrogen chain is closer to a product of local bonds, which localized orbitals express
compactly; this points to localized or natural-orbital pools for spread states.
Since $n_{99}$ of a spread
state grows combinatorially with the number of strongly correlated orbitals, this last
regime belongs to fragment methods such as LAS~\cite{hermes2019lasscf,wang2025lassqd},
which divide the state into weakly coupled pieces, each compact.

\subsection{Predicting it}\label{sec:res-router}
\begin{figure}[t]
\centering
\includegraphics[width=\columnwidth]{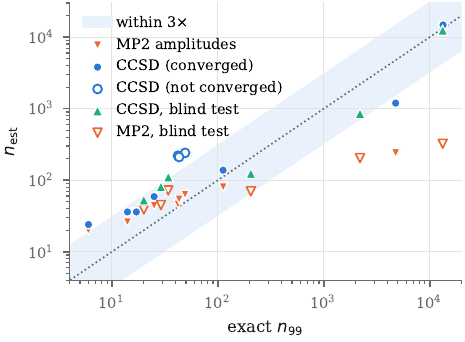}
\caption{Compactness estimate $n_\mathrm{est}$, Eq.~\eqref{eq:nest}, from MP2 and CCSD
amplitudes against the exact $n_{99}$ (MO basis); triangles: blind-test systems. Open symbols: CCSD did not converge (stretched
N$_2$), and the router uses MP2. Shaded band: within a factor of three.}
\label{fig:router}
\end{figure}
Figure~\ref{fig:router} compares $n_\mathrm{est}$ with the exact $n_{99}$. Converged CCSD
amplitudes track $n_{99}$ over more than three orders of magnitude, from 6 (H$_{10}$,
1.0\,\AA) to 13\,376 (H$_{14}$, 2.0\,\AA): within a factor of three for eight of the ten
embeddings used to build the router (and 0.4--3.3 for the six blind-test embeddings,
Sec.~\ref{sec:res-blind}), with H$_{10}$ at 1.0\,\AA\ overestimated four-fold (24 vs.\ 6) and H$_{14}$ at
1.5\,\AA\ underestimated four-fold (1\,195 vs.\ 4\,769). MP2 underestimates spread states by
up to a factor of 40 (H$_{14}$ at 2.0\,\AA: 319), while unconverged CCSD overestimates
stretched N$_2$ four- to five-fold; using CCSD when it converges and MP2 otherwise gives the
best of both. An earlier estimate based on fractionally occupied natural orbitals, which counts
\emph{which} orbitals are correlated, failed by more than an order of magnitude in both
directions (App.~\ref{app:router}): what matters is how many determinants carry weight.

Replayed on all runs, the router of Sec.~\ref{sec:m-router} never recommends \genie\
where it then failed, apart from seed failures (Table~\ref{tab:replay}): \genie\ where it
matched CIPSI, ``deeper'' for N$_2$ at budget 100, and ``not \genie'' for H$_{14}$ at
2.0\,\AA\ with 80 operators, where 160 operators then caught up with CIPSI. Its three
deviations are all on the cautious side: ``deeper'' for H$_{14}$ at 1.0\,\AA\ with budget
200, and ``not \genie'' for the 36-determinant H$_{10}$ embeddings at 1.5 and 2.0\,\AA\ with
$N=20$, where $n_\mathrm{est}$ overestimates $n_{99}$ and \genie\ is exact. The diagnostics cost
one CCSD or MP2 calculation of the embedding, negligible next to training.

\subsection{Blind test on new systems}\label{sec:res-blind}
\begin{table*}[t]
\caption{Blind test of the router on systems it was not tuned on: H$_2$O/6-31G (AVAS
O\,2s2p + H\,1s, 18 qubits, symmetric stretch, $R_\mathrm{OH}$ in \AA) and a ring of twelve
hydrogen atoms (STO-3G, six-atom impurity, 18--20 qubits). Predictions use the rule of
Sec.~\ref{sec:m-router} unchanged and depend only on the embedding. Errors in \kcal\ against
the embedding FCI; best classical: CIPSI or ASCI in the MO basis at the same $N$.}
\label{tab:blind}
\begin{adjustbox}{max width=\textwidth}
\begin{tabular}{lrrrrlllrl}
\toprule
system & $R$ & $w_1$ & $n_\mathrm{est}$ & $n_{99}$ & $N$ & router & \genie\ (seeds) & best classical & ratio\\
\midrule
H$_2$O & 0.96 & 0.97 & 53 & 20 & 200 & \genie & 0.32 & 0.28 & 1.2\\
H$_2$O & 0.96 & 0.97 & 53 & 20 & 1\,864--1\,884 & \genie & 0.004 / 0.005 & 0.0001 & 35--52\\
H$_2$O & 1.44 & 0.87 & 81 & 29 & 200 & deeper & 0.71 & 0.58 & 1.2\\
H$_2$O & 1.44 & 0.87 & 81 & 29 & 2\,433--2\,438 & \genie & 0.003 / 0.003 & 0.00004 & 70\\
H$_2$O & 1.92 & 0.62 & 111 & 34 & 200 & deeper & 1.00 & 0.86 & 1.2\\
H$_2$O & 1.92 & 0.62 & 111 & 34 & 2\,352--2\,355 & \genie & 0.004 / 0.002 & 0.00008 & 21--49\\
H$_{12}$ ring & 1.0 & 0.19 & 124 & 205 & 200 & deeper & 29.3 & 29.2 & 1.0\\
H$_{12}$ ring & 1.0 & 0.19 & 124 & 205 & $10^4$ & \genie & 0.27 / 0.31 & 0.005 & 51--59\\
H$_{12}$ ring & 1.5 & 0.11 & 854 & 2\,197 & 200 & not \genie & 44.1 & 41.7 & 1.1\\
H$_{12}$ ring & 1.5 & 0.11 & 854 & 2\,197 & $10^4$ & \genie & 1.12 / 1.13 & 0.32 & 3.4--3.5\\
\quad 160 ops & 1.5 & & & & $10^4$ & & 0.59 & 0.32 & 1.8\\
H$_{12}$ ring & 2.0 & 0.11 & 12\,531 & 13\,258 & 200 & not \genie & 115 & 44.1 & 2.6\\
H$_{12}$ ring & 2.0 & 0.11 & 12\,531 & 13\,258 & $10^4$ & not \genie & 71.5 / 9.3 & 5.1 & 14 / 1.8\\
\quad 160 ops & 2.0 & & & & $10^4$ & & 8.85 & 5.1 & 1.7\\
\bottomrule
\end{tabular}
\end{adjustbox}
\end{table*}
To test whether the compactness criterion and the router generalize, we chose two systems
after the router had been fixed and ran them unchanged (Table~\ref{tab:blind}): symmetric
bond breaking in water, a polyatomic heteroatom molecule, and a ring of twelve hydrogen
atoms. The ring is a $4n$ ring with a near-degenerate frontier pair, so its leading weight is
only 0.19 already at 1.0\,\AA; CCSD nonetheless converges, and its estimate $w_1^\mathrm{est}=0.92$
is far off. The diagnostics were computed before the classical baselines.

The estimate transfers: $n_\mathrm{est}$ is within a factor of 0.4--3.3 of the exact
$n_{99}$ for all six new embeddings (Fig.~\ref{fig:router}, triangles). So does the
classification. Water is compact at every bond length ($n_{99}=20$--34): \genie\ is within
1.2 of CIPSI at 200 determinants, the router's ``deeper'' at 1.44 and 1.92\,\AA\ was again
cautious---and below 0.01\,\kcal\ without cap. The ring at 2.0\,\AA\ ($n_{99}=13\,258$) is
spread like H$_{14}$ at 2.0\,\AA\ ($13\,376$) and behaves like it: the router says ``not
\genie'', one seed fails ($14\times$ behind ASCI, the set missing 45\,\% of the exact weight)
and the other reaches $1.8\times$, while every selector is 5--8\,\kcal\ off. At 1.5\,\AA\
($n_{99}/N=0.22$) the router says \genie, which ends 1.1\,\kcal\ from the exact energy but
3.5 times behind ASCI, the largest gap in the compact regime apart from seed failures.
As for H$_{14}$, a 160-operator circuit closes most of these gaps: with the failing seed at
2.0\,\AA\ the error drops from 71.5 to 8.85\,\kcal\ ($1.7\times$ ASCI), and at 1.5\,\AA\ from
1.12 to 0.59\,\kcal\ ($1.8\times$). In
the ring at 1.0\,\AA, where the leading weight is only 0.19, the state is nevertheless
compact ($n_{99}=205$) and \genie\ reaches 0.3\,\kcal; with $N=200<n_{99}$ every
selector fails alike. Over the twelve blind cases the router never recommended \genie\
where it then failed by more than a factor of 3.5 at the given budget, apart from the
large-$N$ precision floor discussed next; its other misses are on the cautious side.

The large-$N$ rows show a second, distinct shortfall. Where the budget far exceeds
$n_{99}$, CIPSI converges to $10^{-5}$--$10^{-2}$\,\kcal\ while \genie\ stops at
$10^{-3}$--$0.3$\,\kcal, 20--70 times higher (and 40--700 times for N$_2$ and H$_{14}$ at
1.0\,\AA; Fig.~\ref{fig:compact}, open symbols). These errors are far below chemical
accuracy, but they show that the learned circuit does not converge as fast as classical
selection once the important determinants are found: the remaining small-weight
determinants are reached by chance rather than by selection.

\subsection{Scope and the LAS alternative}\label{sec:res-routes}
This is an impurity-level study: every error above is measured against the exact solution
of the same embedding Hamiltonian, which isolates the solver. The total DMET energy of a
single impurity is not a whole-molecule energy; for H$_{10}$ it lies 77--386\,\kcal\
above whole-molecule FCI, increasingly so on stretching, independently of the solver
(App.~\ref{app:routes}). A whole-molecule comparison requires all fragments and an energy
partitioning, which we leave to future work, as do the embedding errors that
impurity-level benchmarks cannot see~\cite{shajan2025dmetsqd,patra2025dmetsqd}.

For spread states the router points to fragment methods. On H$_{10}$, where whole-molecule
FCI is available, LASSQD agrees with LASSCF to within 1\,\kcal, but the error against FCI
depends on the fragmentation: 38.5, 41.0 and 19.5\,\kcal\ with H$_2$-unit fragments at 1.0,
1.5 and 2.0\,\AA, and 84.5, 32.5 and 9.5\,\kcal\ with two halves (App.~\ref{app:routes}). For
[Fe(H$_2$O)$_4$]$_2$bpym$^{4+}$, where a DMET impurity would need about 40 qubits, LAS needs two
ten-qubit fragments; our LASSCF lies 1.05\,m$\Eh$ below the published
value~\cite{wang2025lassqd}, and LASSQD warm-started from it reproduces it to
$0.3\,\mu\Eh$ for all four circuit/seed combinations (App.~\ref{app:fe2}).

\subsection{Resources and hardware feasibility}\label{sec:res-cost}
\begin{figure*}[t]
\centering
\includegraphics[width=\textwidth]{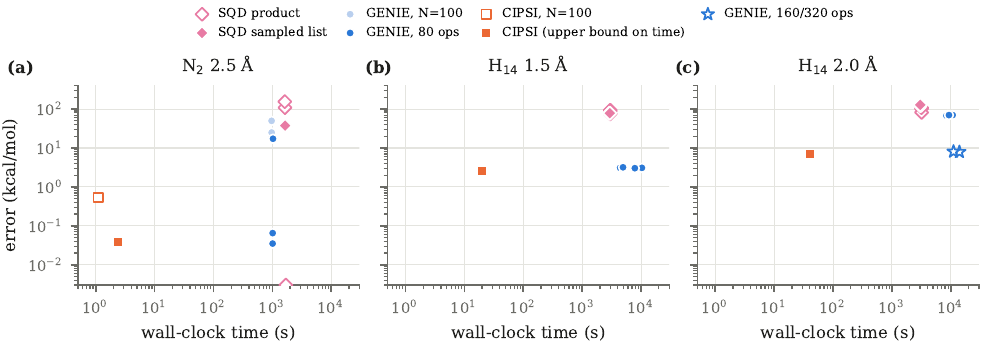}
\caption{Error against wall-clock time per geometry, every run shown. \genie: one training
run on an NVIDIA A100 with 24 CPU threads. DMET+SQD: one run on 10--24 CPU threads plus the
one-time LUCJ optimization of that geometry (1\,555--3\,010\,s). CIPSI: the recorded time also
includes the exact diagonalization and the oracle, so it is an upper bound. At 2.5\,\AA\
(panel a), SQD's product subspace with $10^4$ shots covers the whole space (error 0, drawn
at the axis).}
\label{fig:cost}
\end{figure*}
\begin{table}[t]
\caption{Resources per geometry. Times are wall-clock for one run (one seed); CX counts
are for the best circuit, before routing for \genie\ and after heavy-hex routing for LUCJ.}
\label{tab:resources}
\begin{adjustbox}{max width=\columnwidth}
\begin{tabular}{lllll}
\toprule
 & \genie & DMET+SQD & CIPSI / ASCI\\
\midrule
hardware & A100 + 24 threads & 10--24 threads & 1 node\\
time, N$_2$ (18--20 qb) & 700--5\,100\,s & 1\,600--2\,000\,s$^a$ & $\le 1$--12\,s\\
time, H$_{14}$ (20 qb) & 0.3--3.9\,h & 0.8--0.9\,h$^a$ & $\le 17$--44\,s\\
time, N$_2$/cc-pVDZ (24--26 qb) & 0.8--2.3\,h & 0.03--1.3\,h & $\le 0.1$--0.6\,h$^b$\\
circuits executed & 12\,000 & 1 & 0\\
shots & $1.2\times10^9$ & $10^4$--$10^5$ & 0\\
CX per circuit & 4.4--46\,k & 3\,499$^c$ & --\\
classical diagonalizations & ${\sim}12\,000$ + refinement & 6 & $\sim\log_2 N$\\
\bottomrule
\end{tabular}
\end{adjustbox}
\par\raggedright\footnotesize $^a$Including the one-time LUCJ optimization (25--50\,min).
$^b$Including exact diagonalization in $2.9\times10^6$ determinants.
$^c$All-pairs LUCJ, two repetitions, after heavy-hex routing (N$_2$, 20 qubits).
\end{table}
Figure~\ref{fig:cost} and Table~\ref{tab:resources} compare the resources. At equal
subspace size \genie\ is the most expensive method on every count. It takes 200--600 times
longer than CIPSI in wall-clock time for N$_2$ and H$_{14}$ (4--20 times for cc-pVDZ, where the
recorded CIPSI time is dominated by the exact reference) while not being more accurate, executes $10^4$ times more
circuits than SQD, and its circuits are 5--40 times deeper. Against SQD it buys accuracy:
for H$_{14}$ at 1.5\,\AA\ SQD is 75--94\,\kcal\ off after 0.8\,h, \genie\ 3.0--3.2\,\kcal\
after 1.2--2.9\,h. Against classical selection it buys nothing in these systems; the
comparison with CIPSI is lost by two to three orders of magnitude in time.

The circuits are not executable on present hardware. With an effective error $p$ per CX
the circuit fidelity is roughly $e^{-p\,n_\mathrm{CX}}$. At the error rates of our noise
model ($p\approx8.5\times10^{-3}$, which reproduces the estimated fidelity $1.4\times10^{-13}$
of the routed 3\,499-CX LUCJ circuit), the 80-operator N$_2$ circuits (7\,400 CX) have a
fidelity of $10^{-27}$ and the H$_{14}$ circuits below $10^{-39}$. Under such noise the measured
bitstrings are close to uniform, and, as shown for SQD (App.~\ref{app:noise}), the subspace
is then filled rather than selected; no learned structure survives. The fidelity reaches
0.35--0.5 for the 80-operator circuits only at $p=10^{-4}$, and 0.6--0.9 for all circuits at
$p=10^{-5}$, i.e.\ with error-corrected logical gates. Training would also have to run on
the device, adding $1.2\times10^9$ shots per geometry. \genie\ in its present form
therefore targets early fault-tolerant hardware. Shorter circuits are the obvious lever:
gate-penalized rewards and gate-efficient pools reduced two-qubit counts by 83--98\,\% in
Ref.~\cite{kemmoku2026gqeqsci}.

\section{Conclusion and discussion}\label{sec:conclusion}
Within DMET, \genie\ is, to our knowledge, the first quantum impurity solver whose sampling circuit is
learned. Measured at equal subspace size against classical selection rather than against
the embedded FCI, it closes the gap of one to three orders of magnitude between fixed-form
sampling and classical selection. It reaches CIPSI quality at small budgets but not ASCI
quality, and in a 26-qubit embedding it stays an order of magnitude behind CIPSI at its own
uncapped size, at two to three orders of magnitude more wall-clock time and with circuits of
$10^4$ CX gates that require fault-tolerant hardware. Standard DMET+SQD reaches good energies
by filling the space, which noise makes complete; where filling is not possible, as for H$_{14}$, it fails by
75--130\,\kcal. Both
observations are consistent with critical assessments of
QSCI~\cite{reinholdt2025critical,mlsqdreview2026} and with the strength of modern classical
SCI~\cite{zhang2025trimci}.

The compactness criterion held in a blind test on two new systems, where it predicted
the failure for the stretched hydrogen ring and success for water. It is not specific to
\genie. It applies to any quantum selection
scheme, including Krylov variants~\cite{skqd2025,sqdrift2025} and DMET with learned
classical post-processing~\cite{patra2026rbm}: a sampler can pay off only where the state is
compact in some representation and yet hard to find classically, a regime none of our
systems reaches, and the one in which claimed advantages have so far been
shown~\cite{kirby2026sparse}. For spread states, localized orbitals, in which stretched
H$_{14}$ is 1.3--1.8 times more compact, paired or generalized excitation pools and
multireference initial states are the natural next steps; on
noisy hardware, configuration recovery~\cite{robledo2025sqd} becomes relevant to \genie\
as well.

Methodologically we recommend that quantum subspace methods report the subspace size and
compare against CIPSI and ASCI at that size, in the orbital basis in which those methods
perform best; that seed statistics be reported; and that embedding studies separate the
solver error, as here, from the embedding error against whole-molecule references. A cheap compactness estimate should accompany any claim of where a
method is expected to work.

\section{Code availability}\label{sec:code}
The QuEnAIS package (DMET, LAS, \genie\ interface, classical baselines, router) and all
benchmark grids used here are archived as version 0.3.0 at
\href{https://doi.org/10.5281/zenodo.23736693}{doi:10.5281/zenodo.23736693}~\cite{quenais}
(tag \texttt{v0.3.0} of the GitHub repository); \texttt{docs/REPRODUCE\_GENIE\_PAPER.md}
maps every result to its benchmark grid. Benchmark results, exported analysis data,
embedding Hamiltonians and the figure sources are archived at
\href{https://doi.org/10.5281/zenodo.23737678}{doi:10.5281/zenodo.23737678}~\cite{geniedata}.
Third-party components (the GQE-for-QSCI trainer, PyCI, block2) are not redistributed and
are installed from their upstream sources. Reproduction details are given in
App.~\ref{app:comp}.

\begin{acknowledgments}
This work was carried out at the Fraunhofer Institute for Industrial Mathematics ITWM in
the QuEnAIS project. We thank the ITWM department of High Performance Computing for
computing resources and support.
Calculations were performed on the ITWM computing cluster (NVIDIA A100 GPU nodes and CPU
nodes).
\end{acknowledgments}

\bibliography{references}

\clearpage
\appendix

\section{Systems and DMET implementation}\label{app:dmet}
The impurity is selected with AVAS~\cite{sayfutyarova2017avas} on atom-indexed labels. The
impurity orbitals and their Schmidt bath, obtained from the reference one-particle density
matrix, span the embedding space. In detail:
(i)~\emph{Mean field.} The whole molecule is described by UHF. For N$_2$, H$_{10}$, H$_{14}$,
H$_2$O and the H$_{12}$ ring at 2.0\,\AA\ it coincides with RHF to $<10^{-5}$\,\kcal. For the
H$_{12}$ ring at 1.0 and 1.5\,\AA\ it breaks spin symmetry and lies 53.5 and 5.2\,\kcal\
below RHF, as expected for a $4n$ ring with a near-degenerate frontier pair. Only the
environment and $E_\mathrm{core}$ depend on this choice; all solver errors are measured
against the exact solution of the same embedding Hamiltonian, and the embedded reference
for the circuits is the closed-shell embedded RHF.
(ii)~\emph{Impurity.} The impurity orbitals are the AVAS active orbitals (molecular
orbitals of mixed atomic character, not localized atomic orbitals).
(iii)~\emph{Reference density.} Instead of the mean-field density, the Schmidt decomposition
acts on a CASCI density in the AVAS space, with the active one-electron Hamiltonian screened
by the mean-field potential of the non-active electrons, combined with the mean-field
density outside the active space. This places the bath where the correlated density
couples impurity and environment.
(iv)~\emph{Bath.} The impurity--environment block of the reference density (L\"owdin basis)
is decomposed by singular value decomposition; singular values below $10^{-8}$ are
discarded, so a bath of zero orbitals is allowed. The number of bath orbitals is the larger
of the position of the largest gap in the singular values and the number needed to cover
99.9\,\% of $\sum\sigma^2$, capped at the number of impurity orbitals and at 18 embedding
orbitals in total.
(v)~\emph{Electron count.} The number of embedded electrons is the rounded trace of the
reference density in the embedding space, not the active-space count.
(vi)~\emph{Environment.} The environment density is the reference density projected onto
the complement of the embedding space; it enters $h_{pq}$ through Coulomb and exchange
terms. $E_\mathrm{core}$ is fixed so that the projected UHF density reproduces the
whole-molecule UHF energy within the embedding Hamiltonian.
(vii)~\emph{Chemical potential.} A one-shot grand-canonical shift $h\to h-\mu\mathbb 1$,
$E_\mathrm{core}\to E_\mathrm{core}+\mu N$ is applied; it cancels exactly for the
fixed-particle-number solvers used here.
(viii)~\emph{One-shot.} No correlation potential is fitted and no self-consistency loop is
run; the energy reported is that of the embedding Hamiltonian (Sec.~\ref{sec:res-routes}).
Table~\ref{tab:systems} lists every system, its embedding and the resulting qubit count
(Jordan--Wigner, two qubits per spatial orbital).

\begin{table*}[t]
\caption{Systems, embeddings and problem sizes.}
\label{tab:systems}
\begin{adjustbox}{max width=\textwidth}
\begin{tabular}{llllll}
\toprule
system & basis & $R$ (\AA) & impurity / fragments & embedding (orbitals; $N_\alpha,N_\beta$) & qubits; determinants\\
\midrule
N$_2$ & STO-3G & 1.1--2.5 & AVAS N\,2s2p (10e,8o) & full valence space, 9--10 orbitals & 18--20; $\le 7\,056$\\
N$_2$ & cc-pVDZ & 1.1, 2.1 & AVAS N\,2s2p & 8 impurity + 5 bath = 13 of 28; (7,7) & 26; $2\,944\,656$\\
H$_{10}$ & STO-3G & 1.0, 1.5, 2.0 & 1s of atoms 4, 5 & $\le 6$ of 10 & $\le 12$; small\\
H$_{14}$ & STO-3G & 1.0 & 1s of atoms 5--8 & 9 of 14; (4,4) & 18; 15\,876\\
H$_{14}$ & STO-3G & 1.5, 2.0 & 1s of atoms 5--8 & 10 of 14; (5,5) & 20; 63\,504\\
H$_2$O (blind) & 6-31G & 0.96, 1.44, 1.92 & AVAS O\,2s2p + H\,1s & 9 orbitals; (5,5) & 18; 15\,876\\
H$_{12}$ ring (blind) & STO-3G & 1.0, 1.5, 2.0 & 1s of six adjacent atoms & 9--10 orbitals & 18--20; $\le$63\,504\\
\midrule
H$_{10}$, H$_{14}$ (LAS) & STO-3G & 1.0--2.0 & H$_2$ units (2o, 1$\alpha$1$\beta$) or two halves (5o/7o) & -- & 4 or 10/14 per fragment\\
{}[Fe(H$_2$O)$_4$]$_2$bpym$^{4+}$ & 6-31G & equilibrium & two Fe\,3d: (5o; 4$\alpha$2$\beta$), (5o; 2$\alpha$4$\beta$) & AVAS (12e,10o) & 10 per fragment\\
\bottomrule
\end{tabular}
\end{adjustbox}
\end{table*}

\section{Orbital-basis consistency}\label{app:basis}
The \genie\ operator pool is built from CCSD amplitudes of the embedded RHF and therefore
refers to the embedded MO basis, and the circuit starts from the lowest orbitals occupied.
In the raw Schmidt basis that determinant is not Hartree--Fock: for N$_2$ the overlap
between its occupied space and the RHF occupied space can be as small as $10^{-3}$, and its
energy lies 20\,000--40\,000\,\kcal\ above HF. The circuit then starts in the wrong state,
and the pool's excitation indices address the wrong orbitals. After rotation, energies are
unchanged; determinant indices and weights ($w_1$, oracle ordering) refer to the MO basis.
Table~\ref{tab:basis} shows the effect.

\begin{table}[h]
\caption{N$_2$/STO-3G, \genie\ error (\kcal) before and after the basis rotation (CCSD
angles, 80 operators, no cap).}
\label{tab:basis}
\begin{adjustbox}{max width=\columnwidth}
\begin{tabular}{lcc}
\toprule
$R$ (\AA) & raw basis (3 seeds) & MO basis (2 seeds)\\
\midrule
1.1 & $46\pm0.8$ & 0.04--0.05\\
1.8 & $82\pm0.3$ & 0.005--0.009\\
2.1 & $95\pm0.04$ & 0.0013--0.0015\\
2.5 & $78\pm0.6$ & 0.0003\\
\bottomrule
\end{tabular}
\end{adjustbox}
\end{table}

\section{Classical baselines and full N$_2$ data}\label{app:baselines}
\paragraph*{CIPSI.} Starting from the lowest-diagonal determinant, the set is grown by
adding the candidates with the largest Epstein--Nesbet contributions
$|\langle D_a|H|\Psi\rangle|^2/(E-H_{aa})$, doubling the size per iteration and
re-diagonalizing, until exactly $N$ determinants are reached.

\paragraph*{ASCI.} Seeded with the CIPSI set; each iteration re-ranks members by $|c|$ and
candidates by their first-order amplitude and keeps the best $N$, so determinants can leave
the set. ASCI can beat the oracle because the largest-$|c|$ set is not energy-optimal.

\paragraph*{Fixed-size HCI.} Heat-bath CI adds $a$ if $\max_i |H_{ai}c_i|>\epsilon$. To
compare at fixed size, $\log_{10}\epsilon$ is bisected (30 steps between $-8$ and 1) for
the largest converged HCI set with at most $N$ determinants, which is topped up to exactly
$N$ by heat-bath score in one step and re-diagonalized. The one-step top-up and the absence
of an energy denominator make this variant weaker than CIPSI at small $N$; it is reported
for completeness and is not the reference baseline; production semistochastic HCI with
an energy-based size control would be closer to CIPSI. For spaces above ${\sim}10^5$ determinants HCI is skipped.

\begin{table*}[t]
\caption{N$_2$/STO-3G, all selectors at equal size $N$ (\kcal\ vs.\ embedding FCI).
``raw''/``MO'': selection basis. \genie: CCSD angles, 80 operators, 120 epochs, seeds 4/5/6
(4/5 for the uncapped runs). Bold: best classical.}
\label{tab:n2full}
\begin{adjustbox}{max width=\textwidth}
\begin{tabular}{rrlrrrrrrr}
\toprule
$R$ & $N$ & \genie\ (seeds) & HCI raw & HCI MO & CIPSI raw & CIPSI MO & ASCI raw & ASCI MO & oracle MO\\
\midrule
1.8 & 100 & 22.76 / 21.51 / 21.49 & 17.89 & 8.38 & 10.37 & 2.01 & 7.81 & \textbf{1.76} & 1.92\\
2.1 & 100 & 18.36 / 0.76 / 16.74 & 13.79 & 6.10 & 7.46 & 0.66 & 6.14 & \textbf{0.53} & 0.57\\
2.5 & 100 & 21.14 / 49.93 / 24.98 & 9.89 & 1.56 & 1.04 & 0.54 & 1.01 & \textbf{0.10} & 0.11\\
\midrule
1.8 & 200 & 0.334 / 0.431 / 0.265 & 12.95 & 1.56 & 2.50 & 0.275 & 2.29 & \textbf{0.245} & 0.253\\
2.1 & 200 & 0.099 / 0.092 / 0.078 & 10.95 & 5.24 & 2.95 & 0.081 & 2.05 & \textbf{0.057} & 0.058\\
2.5 & 200 & 0.065 / 17.31 / 0.035 & 9.64 & 1.32 & 0.63 & 0.039 & 0.36 & \textbf{0.012} & 0.015\\
\midrule
1.1 & 1\,109 / 1\,065 & 0.041 / 0.046 & 0.011 & 0.0002 & 0.002 & 0.0001 & 0.002 & \textbf{0.0001} & 0.0001\\
1.8 & 964 / 1\,035 & 0.009 / 0.005 & 0.051 & 0.0002 & 0.003 & 0.0001 & 0.002 & \textbf{0.0001} & 0.0001\\
2.1 & 660 / 672 & 0.0013 / 0.0015 & 0.080 & 0.0000 & 0.003 & 0.0000 & 0.003 & \textbf{0.0000} & 0.0000\\
2.5 & 586 / 601 & 0.0003 / 0.0003 & 0.175 & $-$0.0001$^a$ & 0.002 & $-$0.0001$^a$ & 0.001 & $-$0.0001$^a$ & $-$0.0001$^a$\\
\bottomrule
\end{tabular}
\end{adjustbox}
\par\raggedright\footnotesize $^a$Numerical noise of the reference ($<2\times10^{-7}\,\Eh$).
Free-angle \genie\ runs reach the embedding FCI with 25--51\,\% of the space at every geometry.
\end{table*}

\paragraph*{Training variance.} In a plain CAS(10e,8o) N$_2$ problem at 2.1\,\AA\
($w_1=0.19$, budget 200, 120 epochs), five seeds gave 47.37, 5.28, 5.11, 5.27 and
0.155\,m$\Eh$ (CIPSI 1.64, oracle 0.110\,m$\Eh$). An earlier ``\genie\ beats CIPSI'' claim
based on a single favorable run was withdrawn on this basis.

\section{DMET with SQD}\label{app:sqd}
\paragraph*{Circuit.} Heavy-hex pair patterns lose most of the correlation even after
optimization (Table~\ref{tab:lucj}); we therefore use all-pairs LUCJ with two repetitions,
CCSD-initialized and refined with ten linear-method iterations. One batch containing all
samples and six configuration-recovery iterations are used.

\begin{table}[h]
\caption{H$_4$ (one two-orbital fragment, first LAS cycle; exact off-HF weight 0.0262):
LUCJ variants. Error of the circuit energy (m$\Eh$).}
\label{tab:lucj}
\begin{adjustbox}{max width=\columnwidth}
\begin{tabular}{lcc}
\toprule
LUCJ variant & off-HF weight & $E-E_\mathrm{FCI}$\\
\midrule
heavy-hex, 1 rep, not optimized & 0.0004 & 29.5\\
all pairs, 4 reps, not optimized & 0.0267 & 0.004\\
any variant, optimized & 0.0262 & 0.000\\
\bottomrule
\end{tabular}
\end{adjustbox}
\end{table}

\begin{table*}[t]
\caption{N$_2$/STO-3G, DMET+SQD with optimized LUCJ ($10^4$/$10^5$ shots, seeds 1--3; \kcal\
vs.\ embedding FCI). ``CIPSI @ dim'': CIPSI (raw basis) at the size of the sampled list;
$\langle S^2\rangle$ of the list state.}
\label{tab:sqdlist}
\begin{adjustbox}{max width=\textwidth}
\begin{tabular}{rrrrrrrr}
\toprule
$R$ & shots & product SQD & list SQD & list dim (sampled) & CIPSI @ dim & list/CIPSI & $\langle S^2\rangle$\\
\midrule
1.1 & $10^4$ & 1.2--3.4 & 6.3 (5.6--6.6) & 81 (62) & 4.5 & 1.4 & 0.01\\
1.1 & $10^5$ & 0.27--0.53 & 1.6 (1.5--1.8) & 248 (175) & 0.48 & 3.4 & 0.001\\
1.8 & $10^4$ & 14.7--27.5 & 13.6 (12.7--14.1) & 197 (154) & 2.6 & 5 & 1.2\\
1.8 & $10^5$ & 2.5--9.9 & 10.9 (9.9--11.6) & 361 (294) & 0.56 & 19 & 0.5--1.1\\
2.5 & $10^4$ & 0.5--155 & 37.4 (37.1--37.6) & 139 (107) & 0.80 & 47 & 2.4\\
2.5 & $10^5$ & 0.22--0.29 & 36.4 (35.9--37.0) & 309 (221) & 0.37 & 99 & 2.4--2.6\\
\bottomrule
\end{tabular}
\end{adjustbox}
\end{table*}

\paragraph*{Exploration shots.} Replacing 25--50\,\% of the shots by samples of a random
shallow circuit makes every tested system exact (H$_8$, N$_2$/STO-3G and N$_2$/cc-pVDZ),
with subspaces of 99.9--100\,\% of the space and 30\,min to 3\,h of classical
diagonalization per run.

\begin{table*}[t]
\caption{N$_2$/cc-pVDZ (26 qubits). DMET+SQD with an unoptimized LUCJ circuit (all pairs, 2
reps; optimization would need ${\sim}74$\,GB), seed 1, and \genie\ (seed 4). \kcal\ vs.\
embedding FCI; subspace as fraction of the space ($2.94\times10^6$ at 1.1\,\AA, $8.5\times10^5$ at
2.1\,\AA, where the bath is smaller).}
\label{tab:dz}
\begin{adjustbox}{max width=\textwidth}
\begin{tabular}{rlrr}
\toprule
$R$ & method & error & subspace\\
\midrule
1.1 & SQD $10^4$ / $10^5$ shots & 28.9 / 19.7 & 0.005 / 0.01\,\%\\
1.1 & CIPSI (raw) at those sizes & 5.4 / 1.5 & \\
1.1 & \genie, budget 200 / no cap & 6.25 / 0.48 & 0.007 / 0.34\,\%\\
2.1 & SQD $10^4$ / $10^5$ shots & 94.1 / 88.5 & 0.02 / 0.03\,\%\\
2.1 & CIPSI (raw) at those sizes & 11.2 / 7.9 & \\
2.1 & \genie, budget 200 / no cap & 22.5 / 0.62 & 0.023 / 1.2\,\%\\
1.1 & CIPSI / ASCI (MO) at \genie\ sizes & 6.16, 0.021 / 6.14, 0.021 & \\
2.1 & CIPSI / ASCI (MO) at \genie\ sizes & 17.2, 0.053 / 11.7, 0.053 & \\
\bottomrule
\end{tabular}
\end{adjustbox}
\end{table*}

\section{Noise model}\label{app:noise}
Samples are drawn with Qiskit Aer~\cite{qiskit2024} from a noise model built on the
FakeSherbrooke backend: circuits are routed onto the heavy-hex coupling map, depolarizing
errors with the device-average one- and two-qubit rates multiplied by a scale
$s\in\{0,0.1,1\}$ are attached to every gate, and readout errors are included. The
estimated circuit fidelity is $\prod_g(1-s\,p_g)$; $s=0$ keeps the routing without errors
and must reproduce the noiseless result.

\begin{table*}[t]
\caption{N$_2$/STO-3G, 1.1\,\AA, DMET+SQD under noise (all-pairs LUCJ, 2 repetitions,
$10^4$ shots, seeds 1--3; subspace size, fraction of the 14\,400-determinant space, error in
\kcal). Valid shots: fraction of bitstrings with the correct electron numbers.}
\label{tab:noise}
\begin{adjustbox}{max width=\textwidth}
\begin{tabular}{lrrll}
\toprule
$s$ & fidelity & valid shots & product & list\\
\midrule
0 & 1 & 100\,\% & 676--729 (5\,\%): 1.7--2.9 & 87--95 (0.6\,\%): 5.5--6.7\\
0.1 & 0.052 & 13\,\% & 14\,400 (100\,\%): 0.000 & 5\,570--5\,639 (39\,\%): 0.05--0.20\\
1 & $1.4\times10^{-13}$ & 1.5\,\% & 14\,400 (100\,\%): 0.000 & 7\,550--7\,630 (53\,\%): 0.21--2.0\\
\bottomrule
\end{tabular}
\end{adjustbox}
\end{table*}

\section{Compactness diagnostics and router}\label{app:router}
\paragraph*{HF weight from CCSD.} With closed-shell amplitudes in intermediate
normalization and $\tau_{ij}^{ab}=t_{ij}^{ab}+t_i^at_j^b$, the spin-adapted norms of the
singly and doubly excited parts are
\begin{align}
\|C_1\|^2&=2\sum_{ia}(t_i^a)^2,\\
\|C_2\|^2&=\sum_{ijab}\tau_{ij}^{ab}\left(2\tau_{ij}^{ab}-\tau_{ij}^{ba}\right),
\end{align}
and $w_1^\mathrm{est}=1/(1+\|C_1\|^2+\|C_2\|^2)$. For two electrons in two orbitals this
gives $1/(1+t^2)$, the exact intermediate-normalization weight.

\paragraph*{Determinant-level coefficients.} Each $t_i^a$ represents an $\alpha$ and a
$\beta$ single; each $\tau_{ij}^{ab}$ an $\alpha\beta$ double; for $i<j$, $a<b$ the
antisymmetrized $\tau_{ij}^{ab}-\tau_{ij}^{ba}$ represents an $\alpha\alpha$ and a
$\beta\beta$ double. Overlapping pairs in the quadruple term of Eq.~\eqref{eq:nest} are not
excluded, so this term is an upper estimate. The count is evaluated in $O(n\log n)$ by
sorting $|c_D|$ and counting partners above $\tau/|c_D|$ by binary search, and is capped at
the size of the space.

\paragraph*{A rejected estimate.} Counting natural orbitals with $\min(n,2-n)>0.1$ and
setting $d_\mathrm{est}=\binom{n_f}{n_f/2}^2$ misjudged both families (stretched N$_2$: 4
vs.\ $n_{99}=49$; H$_{14}$ at 1.5\,\AA: 63\,504 vs.\ 4\,769).

\paragraph*{Exact references.} $n_{99}$ is evaluated in the MO basis in which \genie\
works (exported with \texttt{tools/paper\_data.py}); the raw Schmidt-basis value and the
inverse participation ratio $1/\sum_I|c_I|^4$ are listed for comparison
(Table~\ref{tab:diag}). Leading weights agree in both bases to $10^{-3}$.

\begin{table*}[t]
\caption{Diagnostics for every embedding with a \genie\ run.}
\label{tab:diag}
\begin{adjustbox}{max width=\textwidth}
\begin{tabular}{lrrrcrrrrrr}
\toprule
system & $R$ & $w_1^\mathrm{est}$ & $w_1$ (exact) & CCSD conv. & $n_\mathrm{est}$ (MP2) & $n_\mathrm{est}$ (CCSD) & $n_\mathrm{est}$ (used) & $n_{99}$ (MO) & $n_{99}$ (Schmidt) & IPR (Schmidt)\\
\midrule
H$_{10}$ & 1.0 & 0.956 & 0.956 & yes & 20 & 24 & 24 & 6 & 8 & 1.09\\
H$_{10}$ & 1.5 & 0.094$^b$ & 0.840 & yes & 26 & 36 & 36 & 14 & 10 & 1.40\\
H$_{10}$ & 2.0 & 0.631 & 0.628 & yes & 36 & 36 & 36 & 17 & 13 & 2.27\\
H$_{14}$ & 1.0 & 0.899 & 0.894 & yes & 80 & 138 & 138 & 112 & 87 & 1.25\\
H$_{14}$ & 1.5 & 0.460 & 0.458 & yes & 240 & 1\,195 & 1\,195 & 4\,769 & 2\,981 & 4.57\\
H$_{14}$ & 2.0 & 0.189 & 0.144 & yes & 319 & 14\,800 & 14\,800 & 13\,376 & 7\,648 & 37.4\\
N$_2$ & 1.1 & 0.783 & 0.917 & yes & 44 & 59 & 59 & 25 & 33 & 1.19\\
N$_2$ & 1.8 & 0.246 & 0.426 & no & 63 & 241 & 63 & 49 & 90 & 5.11\\
N$_2$ & 2.1 & 0.378 & 0.192 & no & 46 & 220 & 46 & 42 & 91 & 17.3\\
N$_2$ & 2.5 & 0.290 & 0.099 & no & 54 & 211 & 54 & 43 & 64 & 22.2\\
\bottomrule
\end{tabular}
\end{adjustbox}
\par\raggedright\footnotesize $^b$CCSD amplitude blow-up; the weight estimate is unreliable
here, one reason it is not used for routing.
\end{table*}

\begin{table*}[t]
\caption{Router recommendation versus outcome (\genie/best classical at the same $N$; 80
operators unless stated).}
\label{tab:replay}
\begin{adjustbox}{max width=\textwidth}
\begin{tabular}{lrrll}
\toprule
case & $n_\mathrm{est}$ & $N$ & router & outcome\\
\midrule
N$_2$ 1.1 & 59 & 1\,100 & \genie & both $<0.05$\,\kcal\\
N$_2$ 1.8--2.5 & 46--63 & 200 & \genie & 1.1--1.8$^c$\\
N$_2$ 1.8--2.5 & 46--63 & 100 & deeper & 1.4--490\\
H$_{14}$ 1.0 & 138 & $\sim$7\,000 & \genie & $\le0.025$\,\kcal; free angles exact\\
H$_{14}$ 1.0 & 138 & 200 & deeper & 1.02\\
H$_{14}$ 1.5 & 1\,195 & $10^4$ & \genie & 1.2\\
H$_{14}$ 1.5 & 1\,195 & 200 & not \genie & all fail\\
H$_{14}$ 2.0 & 14\,800 & $10^4$ & not \genie & 12 (1.3 at 160 ops)\\
H$_{14}$ 2.0 & 14\,800 & 200 & not \genie & all fail\\
H$_{10}$ 1.0 & 24 & 113 & \genie & exact\\
H$_{10}$ 1.5, 2.0 & 36 & 20 & not \genie & exact$^d$\\
\bottomrule
\end{tabular}
\end{adjustbox}
\par\raggedright\footnotesize $^c$One failed seed at 2.5\,\AA\ (ratio 1\,486); ratios on
errors below 0.1\,\kcal\ (2.1, 2.5\,\AA) reach 1.4--5.6. $^d$36-determinant spaces;
$n_\mathrm{est}=36$ overestimates $n_{99}=14$--17.
\end{table*}

\section{H$_{14}$: full data}\label{app:h14}
Table~\ref{tab:h14full} lists every \genie\ run on H$_{14}$ together with the classical
selectors at the same size, and Table~\ref{tab:h14sqd} the DMET+SQD runs on the same
embeddings.

\begin{table*}[t]
\caption{H$_{14}$, all \genie\ runs and classical selectors at equal size (\kcal\ vs.\
embedding FCI; MO basis; HCI skipped, App.~\ref{app:baselines}).}
\label{tab:h14full}
\begin{adjustbox}{max width=\textwidth}
\begin{tabular}{rrrlrlrrr}
\toprule
$R$ & $N$ & $n_g$ & angles & seed & \genie & CIPSI & ASCI & oracle\\
\midrule
1.0 & 6\,538 & 80 & CCSD & 4 & 0.019 & 0.0004 & 0.0003 & 0.0003\\
1.0 & 6\,273 & 80 & CCSD & 5 & 0.025 & 0.0007 & 0.0006 & 0.0006\\
1.0 & 7\,956 & 80 & free & 4, 5 & 0.000 & 0.000 & 0.000 & 0.000\\
1.0 & 200 & 80 & CCSD & 4 & 4.60 & 4.50 & 4.50 & 4.53\\
1.5 & 10\,000 & 80 & CCSD & 4 / 5 & 3.10 / 3.19 & 2.55 & 2.50 & 2.65\\
1.5 & 10\,000 & 80 & free & 4 / 5 & 3.04 / 3.09 & 2.55 & 2.50 & 2.65\\
1.5 & 10\,000 & 160 & CCSD & 4 & 3.03 & 2.55 & 2.50 & 2.65\\
1.5 & 200 & 80 & CCSD & 4 & 60.5 & 52.2 & 51.8 & 60.7\\
2.0 & 10\,000 & 80 & CCSD & 4 / 5 & 70.1 / 67.9 & 7.00 & 5.95 & 8.76\\
2.0 & 10\,000 & 80 & free & 4 / 5 & 70.2 / 70.2 & 7.00 & 5.95 & 8.76\\
2.0 & 10\,000 & 160 & CCSD & 4 & 7.92 & 7.00 & 5.95 & 8.76\\
2.0 & 10\,000 & 320 & CCSD & 4 & 8.09 & 7.00 & 5.95 & 8.76\\
2.0 & 200 & 80 & CCSD & 4 & 136.0 & 44.5 & 44.4 & 113.1\\
\bottomrule
\end{tabular}
\end{adjustbox}
\par\raggedright\footnotesize All runs 120 epochs. Runs interrupted before 120 epochs are
rejected by the pipeline (two such runs, 4 and 32 epochs, were rerun).
\end{table*}

\begin{table}[h]
\caption{H$_{14}$, DMET+SQD on the same embeddings (optimized LUCJ, $10^4$ shots, seeds
1--3): subspace size, fraction of the space and error (\kcal\ vs.\ embedding FCI).}
\label{tab:h14sqd}
\begin{adjustbox}{max width=\columnwidth}
\begin{tabular}{rlrrr}
\toprule
$R$ & subspace & size & fraction & error\\
\midrule
1.0 & list & 238--267 & 1.5--1.7\,\% & 5.4--6.1\\
1.0 & product & 2\,809--3\,969 & 18--25\,\% & 2.3--2.8\\
1.5 & list & 569--589 & 0.9\,\% & 78.4--78.6\\
1.5 & product & 6\,241--7\,569 & 10--12\,\% & 75--94\\
2.0 & list & 4\,018--4\,127 & 6.3--6.5\,\% & 122--128\\
2.0 & product & 42\,025--46\,225 & 66--73\,\% & 82--109\\
\bottomrule
\end{tabular}
\end{adjustbox}
\end{table}

\section{Whole-molecule energies and LAS}\label{app:routes}
\paragraph*{Single-impurity DMET energy.} For H$_{10}$ (impurity: 1s of atoms 4 and 5) the
total DMET energy with an exact impurity solver lies 77, 198 and 386\,\kcal\ above
whole-molecule FCI at 1.0, 1.5 and 2.0\,\AA: one impurity in a mean-field environment does not
describe a stretched chain. This error is the same for every solver, so it does not
affect the impurity-level comparisons of the main text, but it rules out a
whole-molecule ranking of DMET against LAS with a single impurity.

\paragraph*{LAS on hydrogen chains.} LASSCF/LASSQD~\cite{hermes2019lasscf,wang2025lassqd}
with AVAS on all H\,1s and two fragmentations: H$_2$ units (2 orbitals, 1$\alpha$1$\beta$
each) and two halves (5 or 7 orbitals, opposite-$M_S$ doublets); LASSQD with optimized
LUCJ, $10^4$ shots, seeds 1--3. Table~\ref{tab:las} lists errors against whole-molecule
FCI for H$_{10}$ and total energies for H$_{14}$ (no FCI reference). LASSQD reproduces
LASSCF to better than $10^{-3}$\,\kcal\ with H$_2$-unit fragments (4 qubits) in every case.
With halves it does so in most runs, but lies 1.7--2.7\,\kcal\ above LASSCF for the
seven-orbital (14-qubit) fragments of H$_{14}$ at 1.0\,\AA, and one seed each lies 0.9
(H$_{10}$, 1.0\,\AA) and 0.34\,\kcal\ (H$_{14}$, 1.5\,\AA) above. For H$_{14}$ halves at
2.0\,\AA, LASSQD warm-started from the converged LASSCF ends 8.0, 32 and 41\,\kcal\ above it
for the three seeds: with seven strongly correlated orbitals per fragment, the SQD fragment
solver no longer reproduces the exact fragment solution.

\begin{table}[h]
\caption{LAS on hydrogen chains (STO-3G). H$_{10}$: error against whole-molecule FCI
(\kcal); H$_{14}$: LASSCF total energy ($\Eh$). LASSQD equals LASSCF within the stated
tolerance.}
\label{tab:las}
\begin{adjustbox}{max width=\columnwidth}
\begin{tabular}{lrrr}
\toprule
 & 1.0\,\AA & 1.5\,\AA & 2.0\,\AA\\
\midrule
H$_{10}$, H$_2$ units & 38.5 & 41.0 & 19.5\\
H$_{10}$, halves & 84.5 & 32.5 & 9.5\\
H$_{14}$, H$_2$ units & $-7.43226$ & $-6.89705$ & $-6.59858$\\
H$_{14}$, halves & $-7.41492$ & $-6.95024$ & $-6.63159$\\
\bottomrule
\end{tabular}
\end{adjustbox}
\end{table}
Halves are better than H$_2$ units for stretched chains and worse near equilibrium; the
fragmentation decides the accuracy.

\paragraph*{N$_2$ with atomic fragments.} One fragment per N (4 orbitals, 5 electrons).
Quartet fragments (opposite $M_S$) lie 1\,320\,\kcal\ above FCI at short range and 6.9 at
2.5\,\AA; doublet fragments 34--270. LASSQD (optimized LUCJ, $10^4$ shots) reproduces
LASSCF to $\lesssim0.1$\,\kcal\ except where seeds land on different LASSCF stationary
points (1.6, 2.0, 2.25\,\AA). Non-parallelity errors (\kcal): CASSCF 0.09, CASCI 0.21,
DMET+SQD 29.7, LASSCF (doublet) 237, CCSD(T) 313, HF 435.

\paragraph*{Determinant carryover (ScF/STO-3G, (8e,9o), fragments Sc(6o) + F(3o)).}
LASSQD error vs.\ LASSCF (\kcal, five seeds) with/without carryover~\cite{wang2025lassqd}:
$10^3$ shots $5.5\pm10.8$ / $10.4\pm9.2$; $3\times10^3$: $0.37\pm0.34$ / $5.0\pm4.6$;
$10^4$: $0.14\pm0.32$ / $2.3\pm3.3$; $3\times10^4$: $0.000$ / $0.60\pm1.33$. The threshold
($10^{-4}$--$10^{-6}$) does not matter.

\section{[Fe(H$_2$O)$_4$]$_2$bpym$^{4+}$}\label{app:fe2}
6-31G, charge $+4$, singlet coupling of two high-spin Fe(II); AVAS Fe\,3d (12e,10o);
fragments Fe$_0$ (5o; 4$\alpha$2$\beta$) and Fe$_1$ (5o; 2$\alpha$4$\beta$); geometry from
Ref.~\cite{hermes2020fe2bpym}. Reference LASSCF $-3655.496377\,\Eh$~\cite{wang2025lassqd}.

\paragraph*{LASSCF.} With exact integrals our two-step optimizer (orbital step at fixed
RDMs, then fragment re-solve) reaches $-3655.4974221\,\Eh$ after the 50-cycle limit
(213\,620\,s) with $\Delta E\approx-2.4\times10^{-7}\,\Eh$ per cycle and orbital gradient
$6.6\times10^{-4}$: the energy is converged to ${\sim}5\,\mu\Eh$, the gradient is not,
because the line search stalls along a soft orbital--CI coupled mode. It lies 1.05\,m$\Eh$
below the reference.

\paragraph*{LASSQD.} Four runs warm-started from this LASSCF at fixed orbitals (heavy-hex
and all-pairs LUCJ, seeds 1 and 2), which isolates the fragment-solver error, give
$-3655.4974223\,\Eh$ in every case, $0.24\,\mu\Eh$ below LASSCF ($1.5\times10^{-4}$\,\kcal): the
ten-qubit fragment SQD solves are exact. A self-consistent run with density fitting was
not completed.

\section{Computational details and reproducibility}\label{app:comp}
\paragraph*{Software and hardware.} QuEnAIS~\cite{quenais} (DMET, LAS, baselines, router),
PySCF~\cite{sun2020pyscf}, the GQE-for-QSCI trainer on CUDA-Q~\cite{cudaq} (adapted to
accept embedding Hamiltonians), Qiskit/Qiskit Aer~\cite{qiskit2024} and ffsim~\cite{ffsim}
(versions: PySCF 2.14.0, CUDA-Q 0.15.1, PyTorch 2.13.0 with PyTorch Lightning 2.6.5,
Qiskit 2.5.1, Qiskit Aer 0.17.2, qiskit-addon-sqd 0.12.1, ffsim 0.0.83, NumPy 2.4.6, SciPy 1.17.1). \genie\ runs on an NVIDIA A100 (40\,GB) with 24 CPU threads; SQD, LAS and
baselines on 10--24-thread CPU nodes.

\begin{table}[h]
\caption{\genie\ settings.}
\label{tab:hyper}
\begin{adjustbox}{max width=\columnwidth}
\begin{tabular}{ll}
\toprule
setting & value\\
\midrule
operator pool & fermionic excitations (DMET pool), JW\\
angles & CCSD amplitudes, or $\{\pm1,\pm0.4,\pm0.1\}$\\
operators per circuit $n_g$ & 80 (160, 320 depth test)\\
epochs & 120\\
samples / batch / buffer / warm-up & 100 / 100 / 100 / 100\\
loss & GRPO, clip $[0.8,1.28]$, + NLL of best sequence\\
inverse temperature $\beta$ & 0.5, +0.02 per step, target variance $10^{-5}$\\
optimizer & AdamW, lr $5\times10^{-6}$, weight decay 0.01\\
QSCI cap $N$ & $10^4$ (100, 200 budget runs)\\
orbital basis & embedded MO\\
seeds & 4, 5, 6\\
\bottomrule
\end{tabular}
\end{adjustbox}
\end{table}

\begin{table}[h]
\caption{Wall times and circuit sizes.}
\label{tab:cost}
\begin{adjustbox}{max width=\columnwidth}
\begin{tabular}{lr}
\toprule
task & cost\\
\midrule
\genie, N$_2$/STO-3G & 12--85 min\\
\genie, H$_{14}$ (80 operators) & 0.3--3.0 h\\
\genie, N$_2$/cc-pVDZ & 0.8--2.3 h\\
\genie\ circuit, H$_{14}$ 2.0\,\AA, 160 / 320 ops & $\sim$2.1 / 4.6$\times10^4$ CX$^f$\\
LUCJ optimization, per geometry (18--20 qubits) & 1\,555--3\,010 s once\\
SQD run after optimization & 70--200 s\\
LUCJ after heavy-hex routing, N$_2$ & 1\,204--3\,499 CX, depth 488--3\,852\\
noisy SQD ($s=1$), N$_2$ & 4.3 h\\
LASSCF Fe$_2$, exact integrals & 59 h\\
\bottomrule
\end{tabular}
\end{adjustbox}
\par\raggedright\footnotesize $^f$As reported by the trainer, before routing.
\end{table}

\paragraph*{Reproducibility.} Every number in this paper is a row of a benchmark grid run
with \texttt{tools/run\_benchmark.py} and collected into \texttt{results.csv}; classical
baselines come from \texttt{tools/selection\_baselines.py} and diagnostics from
\texttt{tools/regime\_diagnostics.py --exact}. Grids: \texttt{n2\_gqe\_mo},
\texttt{n2\_gqe\_budget}, \texttt{n2\_sqd\_list}, \texttt{n2\_noise},
\texttt{n2\_dz\_gqe\_mo}, \texttt{h\_chain\_gqe}, \texttt{h14\_gqe\_depth},
\texttt{h\_chain\_compare} (+\,\texttt{\_h14sqd}, \texttt{\_h14halves\_warm}),
\texttt{fe2bpym\_las}, \texttt{fe2bpym\_lassqd\_warm}. Each run
is identified by a hash of its parameters; its directory holds the input, logs, pickled
intermediate results and a one-line \texttt{row.json}.

\end{document}